\documentstyle[12pt,epsf]{article}
\normalbaselines
\makeatletter
\@addtoreset{equation}{section}
\makeatother

\newcommand{\be}{\begin{equation}}
\newcommand{\ee}{\end{equation}}
\newcommand{\bea}{\begin{eqnarray}}
\newcommand{\eea}{\end{eqnarray}}

\title{New Families of Isospectral Hydrogen-like Potentials}

\bigskip \bigskip

\author{J. Oscar Rosas-Ortiz\thanks{On leave of absence from 
         {\it Departamento de F\'\i sica}, CINVESTAV-IPN, 
         {\it A.P. 14-740, 07000 M\'exico D.F., Mexico}. 
           E-mail orosas@fis.cinvestav.mx}\\
       \small {\it Departamento de F\'\i sica Te\'orica\thanks{electronic 
           address:orosas@klander.fam.cie.uva.es}}\\
       \small {\it Universidad de Valladolid, 47011 Valladolid, 
                                    Spain}}

\bigskip \bigskip

\begin{document}

\maketitle
\date{}

\thispagestyle{empty}

\begin{abstract}
By applying algebraic techniques, we construct a two-parametric family of
strictly isospectral Hydrogen-like potentials, as well as some of its
one-parametric limits. An additional one-parametric almost isospectral
family of Hydrogen-like potentials is also investigated. It is argued that
the construction of a SUSY partner Hamiltonian using a factorization
energy $\delta$ less than the ground state energy of the departure
Hamiltonian is unnecessarily restrictive. 
\end{abstract}

\bigskip\bigskip

{\footnotesize
\begin{description}
\item[Key-Words:] Factorization, Hydrogen atom
\item[PACS:] 03.65.Ge, 03.65.Fd, 03.65.Ca 

\end{description}}

\vfill


\newpage

\baselineskip=18pt
\setcounter{page}{1}

\newpage

There now exists a considerable amount of work relating to exactly
solvable one-dimensional potentials in Quantum Mechanics (QM). The main
interest has been to enlarge the number of analitically solvable
potentials using diverse techniques: the Darboux transformation \cite{1},
the Gelfand-Levitan formalism \cite{2}, the standard and modified
factorizations [3-4], the supersymmetric Quantum Mechanics (SUSY QM) 
\cite{5}, etc. The underlying ideas of most of these procedures have been
summarized in an algebraic scheme in which a first order differential
operator intertwines two different Hamiltonians \cite{6}. In particular,
the generation of one-parametric families of potentials isospectral to the
traditional ones [2-9] can be performed by means of this {\it first order
intertwinning technique}. This method leads also in a natural way to the
factorization of the implied Hamiltonians.

Recently, a generalization of this technique, in which the intertwinning
operator is of second order, has been used to derive a two-parametric
family of potentials isospectral to the harmonic oscillator [10]. This
{\it second order intertwinning technique} (SOIT) is a particular case
with $n=2$ of the {\it n}-order intertwinning technique which appears to
be the best way to introduce the higher order SUSY Quantum Mechanics
\cite{11}. In principle, the latter makes possible to generate {\it
n}-parametric families of Hamiltonians isospectral to a given Hamiltonian
\cite{12,13}. 

In this paper we will generate a 2-parametric family of radial isospectral
Hydrogen-like potentials by means of the SOIT. We shall show also that the
one-parametric family of potentials derived by Fern\'andez \cite{8} can be
recovered from ours. As a final result, we will find a 1-parametric family
of potentials having the same energy levels as the corresponding radial
Hydrogen-like potentials except for the ground state energy level.

The standard procedure to deal with Hydrogen-like potentials in QM reduces
to solve the eigenproblem for a particle in a 1-dimensional effective
potential $V_l(r)= l(l+1)/r^2 - 2/r$, where $l=0,1,2,...$ is the azimuthal
quantum number and $r$ is a dimensionless radial coordinate. By
simplicity, instead of working with the standard radial wavefunctions
$R(r)$, we will work with the functions $\psi (r) \equiv r R(r)$ with an
inner product defined by $\langle \psi , \psi' \rangle \equiv 4 \pi
\int_{0}^{+\infty} \bar \psi (r) \psi' (r) dr$. As it is well known, the
eigenvalues of the radial Hamiltonian $H_l = -d^2/dr^2 + V_l(r)$ (with
fixed $l$) are given by:
\be 
E_n \equiv E_{lk} = - \frac{1}{(l+k)^2}; \quad k=1,2,3,...
\ee
where $l+k=n$.

We are looking for a Hamiltonian $\widetilde H_{l'}= - d^2/dr^2 +
\widetilde V_{l'} (r)$, such that the following intertwinning relationship
is satisfied \cite{6}: 
\be
\widetilde H_{l'} A = A H_l,
\ee
where $A$ is a second order differential operator to be determined 
\be
A \equiv \frac{d^2}{dr^2} + \beta (r) \frac{d}{dr} + \gamma 
(r).
\ee
Equations (2-3) and the explicit form of $H_l$ and $\widetilde H_{l'}$
lead to the following relations between $V_{l}(r), \widetilde V_{l'}(r),
\beta (r)$ and $\gamma (r)$:
\bea
\beta \beta'' - \frac{\beta^2}{2} + \left ( 2 \gamma(r) - 
\beta ' - \frac{\beta^2}{2} \right) \beta^2 +
2c=0,\\
\widetilde V_{l'}(r) = \frac{l(l+1)}{r^2} - \frac 2r+ 2 \beta ' ,\\
2 \gamma (r) = \beta^2 - \beta '-2\frac{l(l+1)}{r^2} + \frac 1r -d,
\eea
where $c$ and $d$ are, in principle, arbitrary constants and the prime
denotes derivative with respect to $r$. The key point becomes now to solve
the non-linear second order differential equation (4) for $\beta(r)$. Let
us first enforce that the operator $A$ includes $a_{l-1} a_l$ as a
particular case, where $a_l = (-d/dr + l/r -1/l)$ is the standard
factorization operator for the radial Hydrogen-like Hamiltonians
\cite{3}. This condition permit us, without a loss of generality, to fix
the constants $c$ and $d$ in (4-6) as
\be
c= \frac{(2l-1)^2}{4 l^4 ( l-1)^4}, \quad d= \frac{1 +
(2l-1)^2}{2 l^2 (l-1)^2}.
\ee
After including (7) in (4-6), it turns out that the general solution to
(4) becomes: 
\be
\beta (r) = \frac{1-2l}{l^2(l-1)^2} \left[
\frac{d}{dr} \log \left( \frac{g_2 (r)}{g_1 (r)} \right)
\right]^{-1}, \quad l=2,3,...,
\ee
\be
g_1(r) =\left\{ 1 - 
\frac{\nu_1}{(2l)!} \left( \frac 2l \right)^{2l+1} 
\int_0^r x^{2l} e^{-2x/l} dx \right\} ,
\ee
\be
g_2(r) = e^{r/l(l-1)} \left[ 1 - \frac{r}{l(l-1)}
\right]  \left\{ 1 + \frac{\nu_2}{(2l-1)!}
\left( \frac {2}{l-1} \right)^{2l-1}  \int_0^r \frac{x^{2l}
e^{-2x/(l-1)}}{[l(l-1) -x]^2} dx \right\},
\ee
where $\nu_1$ and $\nu_2$ are integration constants.

The asymptotic behaviour of $\beta (r)$ is given by $\beta (r) \sim
(1-2l)/l(l-1)$, while it diverges as $(2l-1)/r$, when $r\rightarrow 0^+$.
This suggests to us to write the new potential $\widetilde V_{l'}(r)$ in
(5) as
\be
\widetilde V_{l-2}(r) =V_{l-2} (r) + 2 \alpha'(r), 
\quad l=2,3,...,
\ee
where $l'=l-2$ and $\alpha(r) \equiv \beta(r) + (1-2l)/r$, is an
appropriate function that makes evident the limit $\widetilde V_{l-2}(r) 
\rightarrow V_{l-2} (r)$, when $r \rightarrow + \infty$ and $r \rightarrow
0$. The parameter domain for which $\alpha'(r)$ is free of singularities
is given by $\nu_1, \, \nu_2 \in (-\infty, 1)$;  inside this parameter
region, the new two-parametric family of potentials $\widetilde
V_{l-2}(r)$ has the same singularities as $V_{l-2} (r)$. Furthermore, for
$\nu_1= \nu_2 =0$ we have $\alpha' = 0$ and $\widetilde V_{l-2}(r) =
V_{l-2}(r)$. This means that $V_{l-2}(r)$ not only governs the asymptotic
behaviour of $\widetilde V_{l-2} (r)$ but it is a member of the family
(11).

Now, from equation (2) it becomes apparent that the operator $A$ acting on
the eigenfunctions $\{\psi_{nl}(r)\}$ of $H_l$ provides eigenfunctions
$\widetilde \psi_{n,l-2} (r)\propto A \psi_{nl}(r)$, $l=2,3,...$ of
$\widetilde H_{l-2}$ with eigenvalues $\widetilde E_{lk}= E_{lk}$, i.e.: 
\be
\widetilde \psi_{n,l-2} (r) = \frac{l(l-1) n^2}
{\sqrt{(n^2-l^2)(n^2-l^2+2l-1)}} A \psi_{nl}(r).
\ee

However, the set $\{ \widetilde \psi_{n,l-2}(r)\}$ is not yet complete in
the Hilbert space of square-integrable functions ${\cal H}$. This is clear
if we try to answer the following questions: are there functions
$\widetilde \psi_{l-2, \epsilon}(r)$ orthogonal to all the $\widetilde
\psi_{n,l-2} (r)$? If so, are they eigenfunctions of $\widetilde H_{l-2}$?
In order to answer these questions, let us assume that the set $\{ 
\widetilde \psi_{l-2, \epsilon} \}$ exists, then
\[
\langle \widetilde \psi_{l-2, \epsilon} (r), \widetilde 
\psi_{n,l-2} (r) \rangle \propto 
\langle \widetilde \psi_{l-2, \epsilon} (r), A 
\psi_{nl} (r) \rangle = 
\langle A^{\dagger} \widetilde \psi_{l-2, \epsilon} (r), 
\psi_{nl} (r) \rangle = 0.
\]
Due to the fact that $\{ \psi_{nl}(r)\}$ is a complete set in ${\cal H}$,
the kernel of the second order differential operator $A^{\dagger}$ is a
2-dimensional subspace ${\cal H}_\epsilon$ orthogonal to all the
$\widetilde \psi_{n,l-2} (r)$, $l=2,3,...$ Let us write them as
$\widetilde \psi_{l-2, \epsilon} (r) = c_0 \exp [\int f(x) dx]$, where
$c_0$ is a constant and $f(x)$ is to be determined. The equation
$A^{\dagger} \widetilde \psi_{l-2, \epsilon} (r) = 0$ can be rewritten as:

\be
f'(r) - \beta (r) f(r) + f^2 (r) - \beta'(r) + \gamma(r) = 0.
\ee
This Riccati type differential equation has a general solution given by
\be
f(r) = \frac 1l - \frac lr + \beta(r) + \frac {d}{dr} \ln [
c_1 g_1(r) + c_2 g_2(r)],
\ee
where $g_1(r)$ and $g_2(r)$ are given in (9-10), and $c_1$ and $c_2$ are
constants. The generic kernel element is given by $ \widetilde
\psi_{l-2,\epsilon} (r) = C_0 \widetilde \psi_{l-2,0}(r) + C_{-1}
\widetilde \psi_{l-2,-1}(r)$, where
\be 
\widetilde \psi_{l-2,0}(r) = \sqrt{\left( \frac{1 -
\nu_1}{(2l)!} \right) \left( \frac{2}{l} \right)^{2l+1} 
(2l-1)} \left( \frac{1}{l(l-1)} \right) 
\frac{r^l e^{-r/l} g_2(r)}{W(g_1,g_2)}, \quad l=2,3,...,
\ee
and
\be 
\widetilde \psi_{l-2,-1}(r) = \sqrt{\left( \frac{1-\nu_2}
{2l(2l)!} \right) \left( \frac{2}{l-1} \right)^{2l+1} 
(2l-1)} \left( \frac{1}{l(l-1)} \right) 
\frac{r^l e^{-r/l} g_1(r)}{W(g_2,g_1)}, \quad l=2,3,...,
\ee
are both eigenfunctions of $\widetilde H_{l-2}$ with eigenvalues
$\widetilde E_{l-2,0} = -1/l^2$, $\widetilde E_{l-2,-1} = -1/(l-1)^2$,
$l$=2,3,... respectively. In (15-16), $W(g_1,g_2)= g'_1(r)  g_2(r) -
g_1(r) g'_2(r)$ represents the Wronskian of $g_1$ and $g_2$.

Notice that $\{ \widetilde \psi_{l-2,-1}, \widetilde \psi_{l-2,0},
\widetilde \psi_{n,l-2} , \, l=2,3,...\}$ is now a complete set in ${\cal
H}$, and their elements are eigenfunctions of $\widetilde H_{l-2}$,
$l=2,3,...$, with eigenvalues: 
\be 
\widetilde E_{l-2,k} = -\frac{1}{(l-1)^2}, \, -\frac{1}{l^2}, \, 
- \frac{1}{(l+k)^2}; \quad k=1,2,3,...
\ee

Comparing with (1), one can see that this spectrum is identical to that of
$H_{l-2}$, {\it i.e.}, $E_{l-2,k}= \widetilde E_{l-2,k}$. Hence, the
Hamiltonian $\widetilde H_{l-2}$ is strictly isospectral to $H_{l-2}$, and
because $\widetilde V_{l-2} (r)$ depends of two free parameters $\nu_1$
and $\nu_2$, a new two-parametric family of isospectral Hydrogen-like
potentials has been generated. Some particular cases are worth discussing
in more detail.

Firstly, for $\nu_1=\nu_2=0$, the well known Hydrogen-like potential
$V_{l-2}(r)$ is always recovered because it is a member of the family
$\widetilde V_{l-2} (r)$.

Let us now to take $\nu_1=0$ ($\nu_2=0$), in this case the 2-parametric
family $\widetilde V_{l-2}(r)$ is a new 1-parametric family which becomes
to a Hydrogen-like potential when $\nu_2 \rightarrow 0$ ($\nu_1
\rightarrow 0$). In particular, let $\nu_1=0$, and
\[
\nu_2 = \left( \frac{l-1}{2} \right)^{2l-1}
\frac{(2l-2)!}{\gamma_{l-1}}.
\]
In this case, the function $\alpha (r)$ becomes
\[
\alpha (r) = \frac{1-2l}{l(l-1)} + \frac{r^{2l-2} e^{-2r/(l-1)}}
{ \gamma_{l-1} - \int_{0}^{r} x^{2l-2} e^{-2x/(l-1)} dx}, \quad
l=2,3,..
\]
If we change in (11) $l$ by $l+1$ with $\alpha(r)$ as it is defined above,
we get the following one-parametric family of potentials:
\be
\widetilde V_{l-1} (r) = V_{l-1} (r) + 2 \frac{d}{dr} 
\left\{ \frac{r^{2l} e^{-2r/l}}
{ \gamma_{l} - \int_{0}^{r} x^{2l} e^{-2x/l} dx}
\right\}, \quad l=1,2,...
\ee
which was generated by Fern\'andez in 1984 \cite{8}. Notice that, in the
case when $\gamma_l \rightarrow 1/4$, the family (18) gives the particular
case derived by Abraham and Moses \cite{7}. Moreover, when $\gamma_l
\rightarrow \infty$ ($\nu_2 \rightarrow 0$), we have $\widetilde
V_{l-1}(r) \rightarrow V_{l-1}(r)$, just as we have proposed.

The previous potentials can be seen as deformations of $V_{l-1}(r)$
induced by the second term in (18), which does not change the behaviour of
$V_{l-1}(r)$ at the ends of the interval $[0,\infty)$, but can produce
important modifications inside. In particular, there is the possibility of
creating one additional well in $V_{l-1}(r)$ whose depth and position can
be changed by varying $\gamma_l$. On the other hand, the two-parametric
family of potentials (11) admits also the previous interpretation, but
more freedom is given by the deforming term (it depends on two parameters
instead of just one as in the previous case).  Hence, we get the
possibility of introducing now two wells, one of them with its minimum
placed around the global minimum of $V_{l-2}(r)$ and the other one pushed
further out. As in the previous one-parametric case, the depths and
positions of the two wells can be modified by changing the two parameters
$\nu_1$ and $\nu_2$. This is illustrated in Figure 1, where two members of
the family $\widetilde V_1(r)$ of (11) are depicted, together with the
undeformed potential $V_1(r)$ (dashed line). We have plotted also in
Figure 2 the corresponding probability densities for the two energy levels
$\widetilde E_{1,-1}=-1/4$, $\widetilde E_{1,0}= - 1/9$ with $\nu_1 =
\nu_2 = -10$. For the lowest level the probability has a maximum around
the left well. On the other hand, the first excited state has two maxima,
the highest one centered around the right well, while the lowest one is
situated around the left well. When we go over the higher excited states,
the probabilities resemble more and more the corresponding Hydrogen-like
densities. 

Up to now, departing from the Hydrogen-like potential $V_l(r)$ we have
generated a two-parametric family of solvable potentials $\widetilde
V_{l-2}(r)$, with the same spectrum and singularities as $V_{l-2}(r)$.
Now, as the intertwinning operator $A$ is of second order, it is
interesting to look for its possible factorization in terms of two first
order differential operators $b_1$, and $b_2$, {\it i.e.}:  
\be
A=b_2 b_1, \quad b_j = \frac{d}{dr} + w_j(r), \quad j=1,2.
\ee
This leads to $\gamma (r) = w_2 ' + w_1 w_2$, and $ w_1(r)=\beta (r) -
w_2(r)$, where $\beta(r)$ is given by (8), and $w_2$ has the form given in
(14). Hence, $w_1(r)$ takes the form: 
\be
w_1(r) = \frac lr - \frac 1l -\frac{d}{dr} \ln [c_1 g_1(r) 
+ c_2 g_2(r) ].
\ee
It is now clear that when solving (13), we simultaneously have gotten the
solutions to the equation $A^{\dagger} \widetilde \psi_{l-2, \epsilon}
=0$, as well as the factorizations of the operator $A$. There is a
continuous family of factorizations because when we change the values of
$c_1$ and $c_2$ in (20), we are simultaneously changing the operators
$b_1$ and $b_2$, but maintaining fixed their product $A = b_2 b_1$. 

A first consequence of this factorization arises after rewritting equation
(2) as $\widetilde H_{l-2} b_2 b_1 = b_2 b_1 H_l$. Suppose now that there
is a Hamiltonian $H_{l-1}^*= -d^2/dr^2 + V_{l-1}^*$ such that $H_{l-1}^*
b_1 = b_1 H_l$. Thus $b_2 H_{l-1}^* b_1 = b_2 b_1 H_l$, and we get
$\widetilde H_{l-2} b_2 = b_2 H_{l-1}^*$. Therefore, $H_{l-1}^*$ could be
considered as an intermediate Hamiltonian between $H_l$, and $\widetilde
H_{l-2}$.  Hence, the SOIT can be seen as the iteration of two first order
intertwinning transformations. 

We notice here that from the very beginning we are labeling with the
subindex $l-1$ the intermediate Hamiltonian $H^*$. This is a consequence
of the further calculations leading to a centrifugal term for $V^*$ with
exactly that index. 

In order to ensure that the first order intertwinning relationship
$H_{l-1}^* b_1 = b_1 H_l$ would be satisfied, the functions $w_1$ and
$V_{l-1}^*$ must satisfy some restrictions. The key one becomes the
following Riccati equation: 
\be
-w_1' + (w_1)^2 -V_l + \delta_1 =0,
\ee
complemented with the typical SUSY relationship $V_{l-1}^* = V_l + 2 w_1
'$, where $\delta_1$ is a constant to be determined. We notice also that a
first order intertwinning relationship of the kind $H_{l-1}^* b_1 = b_1
H_l$ leads in a natural way to the factorization of the Hamiltonians $H_l$
and $H_{l-1}^*$: $H_l= b_1^{\dagger} b_1 + \delta_1$, and $H_{l-1}^* = b_1
b_1^{\dagger} + \delta_1$. By a similar argument, the Hamiltonians
$H_{l-1}^*$ and $\widetilde H_{l-2}$ become factorized in terms of $b_2$
and $b_2^\dagger$: $H^* = b_2^{\dagger} b_2 + \delta_2$, and $\widetilde H
= b_2 b_2^{\dagger} + \delta_2$. 

In order to determine the intermediate Hamiltonian $H^*_{l-1}$, we must
find inside the family (20) a member obeying equation (21). In fact, for
$c_1=0$ and $c_2=1$, we have $\delta_1 = -1/(l-1)^2$, $l=2,3,...$. Thus,
the potential $V^*_{l-1}(r)$ can be written as: 
\be
V^*_{l-1}(r) = \frac{l(l-1)}{r^2} - \frac 2r + 2 \left[ 
\frac{(g_2')^2 - g_2 '' g_2}{(g_2)^2} \right].
\ee
The parameter domain for which the family (22)  has the same singularity
as $V_{l-1}(r)$ is given by $\nu_2 \in (1, \infty)$. The eigenfunctions of
$H^*_{l-1}$, $l=2,3,...$ are given by $\psi^*_{l-1,-1} \propto r^l
e^{-r/l}/g_2$, and $\psi^*_{n,l-1} \propto b_1 \psi_{nl}$, with
eigenvalues $E^*_{l-1,-1} = -1/(l-1)^2$, and $E^*_{l-1,k}= E_{lk}$,
$k=1,2,...$, respectively. Notice the unusual absence of the state
corresponding to $E^*_{l-1,0}=-1/l^2$. A direct comparison of the spectra
shows that $V^*_{l-1}(r)$ is almost isospectral to $V_{l-1}$, the
difference resting on the ground state level position. 

The next first order intertwinning transformation gives a different
factorization of $H_{l-1}^*$ and some interesting new results. The absent
energy level $\widetilde E_{l-2,0} = E^*_{l-1,0}$ is added now to the
spectrum of $H^*_{l-1}$ in order to generate $\widetilde H_{l-2}$. But
this means that the factorization energy in this second step is greater
than the ground state energy level of $H_{l-1}^*$, and this naturally
fills the {\it hole} generated by the first factorization. The
eigenfunctions of $\widetilde H_{l-2}$, in terms of those of $H^*_{l-1}$,
$l=2,3,...$, are given by $\{ \widetilde \psi_{l-2, -1} \propto b_2
\psi^*_{l-1,-1}, \, \widetilde \psi_{l-2,0}, \, \widetilde \psi_{n,l-2}
\propto b_2 \psi^*_{n, l-1} \}$, where $\widetilde \psi_{l-2,0} \propto
\exp (- \int w_2(x) dx )$ is the eigenfunction associated to the `missing'
energy level $E^*_{l-1,0}=\widetilde E_{l-2,0}$. A direct calculation
shows that this set of eigenfunctions is the same as that derived by means
of the SOIT.


{\bf Concluding Remarks.} In this paper we have shown that the Second
Order Intertwinning Technique allows one to derive a two-parametric family
of isospectral Hydrogen-like potentials. The iteration of two first order
intertwinning transformations leads to the same results but gives
additional information. Thus, against the standard statement of SUSY QM,
there are cases where a factorization energy greater than the ground state
energy of the departure Hamiltonian leads to a physically acceptable SUSY
partner. A deeper discussion of these first order intertwinning cases will
be given elsewhere \cite{13}. Finally, all the potentials presented here
have the same kind of singularity at a fixed point ($r=0$)  as the initial
Hydrogen-like Hamiltonian. The case when the intertwined potentials are
non-singular has been illustrated by Fern\'andez {\it et.al.} for the
harmonic oscillator \cite{6,10,12}, while the case when the SUSY partner
of the oscillator potential has a movable singularity has been
successfully interpreted \cite{14}. The corresponding problem for a family
of isospectral Hydrogen-like potentials with a different singularity as
those of $V_{l}(r)$ in $r=0$ is open.

\section*{Acknowledgements}

This work is supported by a Postdoctoral CONACyT fellowship (M\'exico) in
the program ``{\it Programa de Estancias Posdoctorales en Instituciones
del extranjero 1997-1998}". The author is indebted to Dr. D. J.
Fern\'andez for enlightenning discussions and suggestions.

\vfill\eject


\newpage

\begin{figure}
\centerline{
\epsfbox{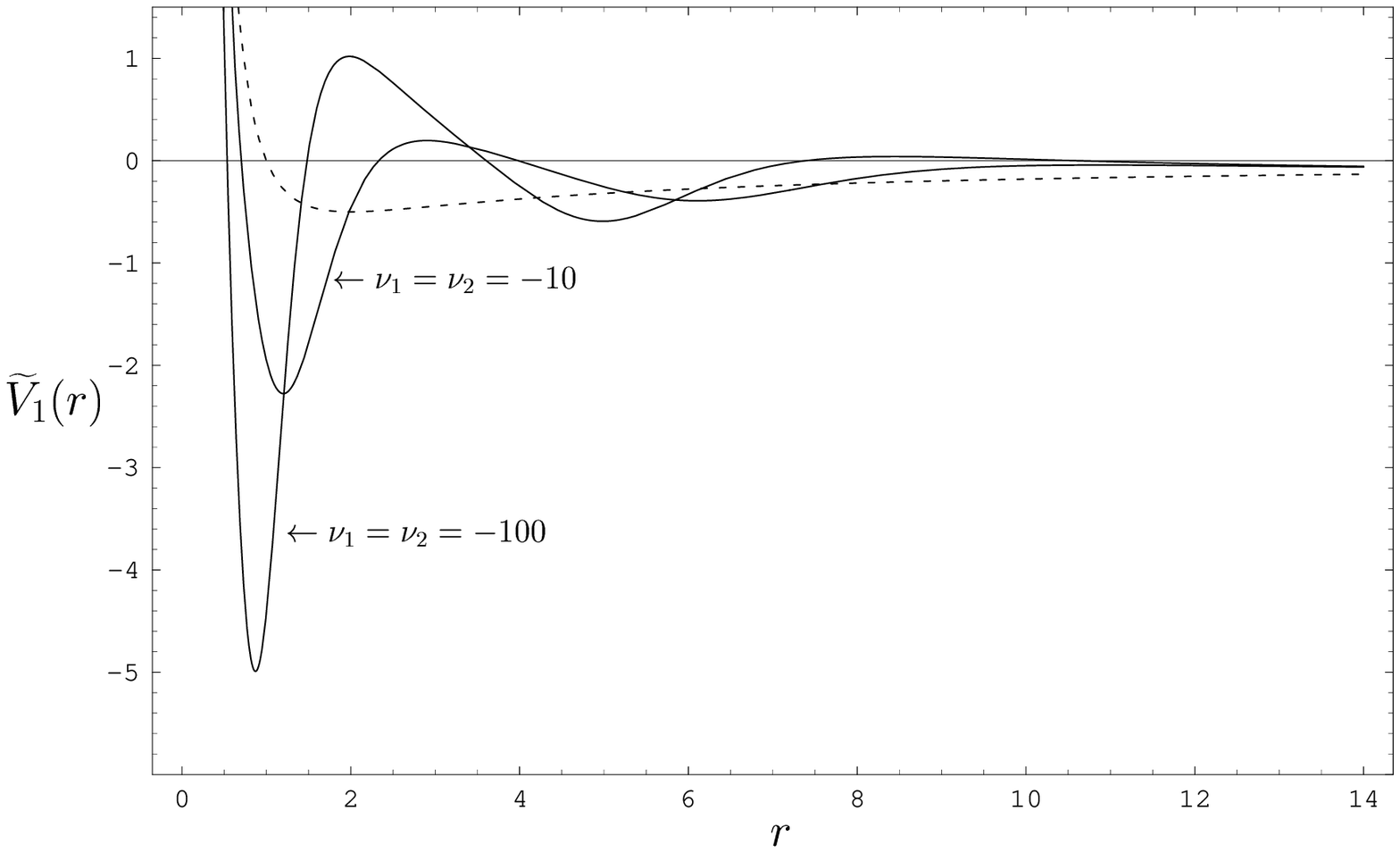}}
\caption{The Hydrogen-like potential $V_{l-2}(r)$ (dashed line) and two
members of the family $\widetilde V_{l-2} (r)$ (solid line) with $l=3$.
Here we have chosen equal values for $\nu_1$ and $\nu_2$, but it is not a
restriction (see eq 11 and bellow). Notice that in the limit case when
$\nu_i \rightarrow 0$, $i=1,2$, the solid lines resemble more and more the
dashed one.}
\end{figure}

\begin{figure}
\centerline{
\epsfbox{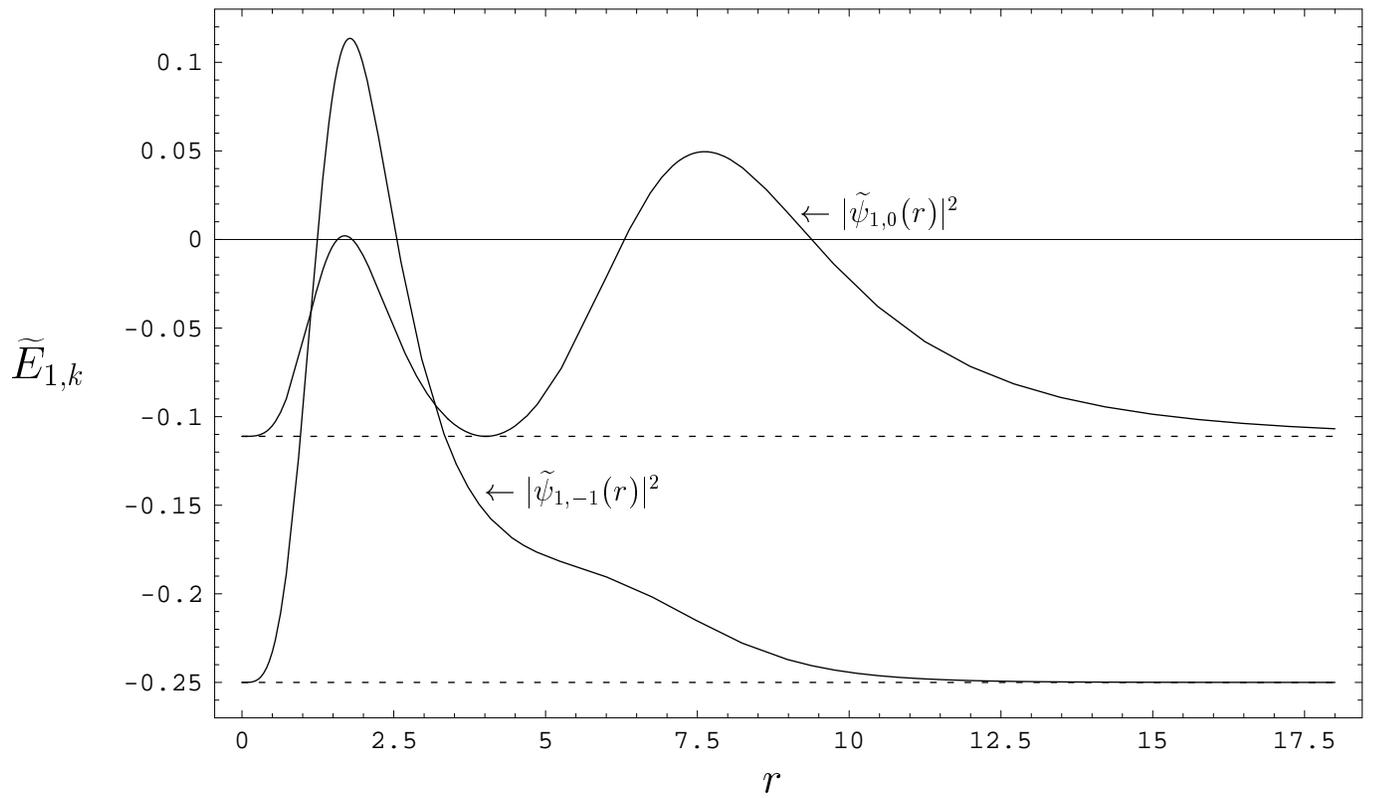}}
\caption{Behaviour of the probability densities of the two first energy 
levels -1/4 and -1/9 of $\widetilde H_1$, with $\nu_1 = \nu_2= -10$. The 
levels are indicated by the dashed lines}
\end{figure}

\end{document}